\documentclass[12pt]{article}
\usepackage{amsmath}
\usepackage{graphicx,psfrag,epsf}
\usepackage{enumerate}
\usepackage{natbib}
\usepackage[linesnumbered,ruled,vlined]{algorithm2e}
\usepackage{adjustbox}
\usepackage{xcolor}
\usepackage{hyperref}

\begin{document}

\title{High Performance Monte Carlo Simulation of Ising Model on TPU Clusters}
\author{Kun Yang, Yi-Fan Chen, George Roumpos, Chris Colby, John Anderson}
\date{\{kuny, yifanchen, roumposg, ccolby, janders\}@google.com}
\maketitle

\begin{abstract}
Large-scale deep learning benefits from an emerging class of AI accelerators. Some of these accelerators' designs are general enough for compute-intensive applications beyond AI and Cloud TPU is one such example. In this paper, we demonstrate a novel approach using TensorFlow on Cloud TPU to simulate the two-dimensional Ising Model. TensorFlow and Cloud TPU framework enable the simple and readable code to express the complicated distributed algorithm without compromising the performance. Our code implementation fits into a small Jupyter Notebook and fully utilizes Cloud TPU's efficient matrix operation and dedicated high speed inter-chip connection. The performance is highly competitive: it outperforms the best published benchmarks to our knowledge by 60\% in single-core and 250\% in multi-core with good linear scaling. When compared to Tesla V100 GPU, the single-core performance maintains a $\sim$10\% gain.  We also demonstrate that using low precision arithmetic---bfloat16---does not compromise the correctness of the simulation results.
\end{abstract}

\section{Introduction}
The Ising model \citep{Ising1925}, which considers short-range interactions between spin variables on the sites of a $d$-dimensional lattice, plays an important role in statistical physics as a prototyping system to study the universal behavior of critical phenomena. Many significant breakthroughs in statistical physics are attributed to the study of the model from either its computational or its theoretical perspective. It is well known that the Ising model has no phase transition in one dimension; however, it undergoes a second-order phase transition between an ordered and a disordered phase in two dimensions or more \citep{Onsager1944, Binder2001}. The critical temperature $T_c$ at which this phase transition occurs on a two-dimensional square lattice was analytically solved by Lars Onsager \citep{Onsager1944}, but it is still an open problem in three or more dimensions. Computer simulation offers a powerful alternative to study such systems and determine critical temperatures, thanks to the development of finite scaling theory \citep{Binder1981} and availability of increasing computational power. This approach ushered in a plethora of interdisciplinary applications outside of physics, including bioinformatics \citep{Barradas-Bautista2018}, economics \citep{Preis2008} and operations research \citep{Cook2018, Prieto-Castrillo2018}.

Large-scale simulation of systems such as Ising model requires a large amount of high performance computing resources, which are usually available in multi-core computing architectures based on distributed shared memory, or distributed clusters (a.k.a data-centers) with homogeneous or heterogeneous nodes commonly seen in private or commercial clouds. Benefiting from the explosion of machine learning, especially deep learning, commercial clouds provide not only CPUs and GPUs, but also specialized chips such as FPGAs and other in-house processors. The Tensor Processing Unit (``Cloud TPU'' or ``TPU'' for short)---an AI application-specific integrated circuit (ASIC) developed by Google for neural network machine learning---has received much attention in the machine learning community \citep{Jouppi2017, Jouppi2019}. Its latest release, Cloud TPU v3,  offers $420\times 10^{12}$ floating-point operations per second (FLOPS) and $128$GB of high bandwidth memory (HBM)\footnote{\raggedright\url{cloud.google.com/tpu/}\label{fn:cloud_tpu_fn}}. Multiple units are connected to form a ``POD'' (Cloud TPU v3 Pod) through a dedicated high speed $2$-D toroidal mesh network, allowing up to $100+$ peta-FLOPS and $32$TB of HBM$^{\ref{fn:cloud_tpu_fn}}$ to be accessed by the application with very low latency and in lockstep. TPU is programmable via software frontends such as TensorFlow \citep{Abadi2016} or PyTorch \citep{Paszke2017}, and can be deployed both for training huge deep neural networks and for performing low-latency online prediction. \citep{Jouppi2017a} reports impressive acceleration of training and online prediction.

With the tremendous amount of computation resources that TPU offers, it is compelling to also consider the opportunities TPU brings for applications beyond machine learning. The programming frontends that are used for TPU, such as TensorFlow, also offer a rich set of functionalities that are highly relevant for scientific computations. The TensorFlow TPU programming stack also provides the additional benefits of allowing distributed algorithms to be expressed with simple and easy-to-understand code without sacrificing performance. In addition, the ability to program conventional scientific simulations in TensorFlow framework makes it easier to explore the hybrid approaches employing both conventional scientific computation methods and modern machine learning techniques on the same framework.

Motivated by these observations, we developed a Single Instruction, Multiple Data (SIMD) distributed Markov Chain Monte Carlo (MCMC) simulation of the two-dimensional Ising model that is programmed in TensorFlow to run on TPU. We demonstrate that such an approach for scientific simulations is very promising. Our code implementation is easy to understand, with entire source code $\sim600$ lines, while also achieves competitive performance and scaling properties when compared with the state-of-the-art benchmarks in terms of both speed and scalability. Another interesting observation is that the lower precision arithmetic---bfloat16 (1 sign bit, 8 exponent bits and 7 mantissa bits \footnote{\raggedright\url{en.wikipedia.org/wiki/Bfloat16_floating-point_format}}) instead of float32---does not compromise the accuracy of the result.

The implementation is open-sourced (URL available in AD Appendix) and can be run through Colaboratory\footnote{https://colab.research.google.com}, a free cloud service based on Jupyter Notebooks for interactive data science. It is also worth pointing out that all results, including large-scale distributed computation, can be run through the Notebook frontend with minimal setup (mainly for allocating TPU backends and setting up the connection to the TPU backend).

In the following sections, we first review the high level architecture of TPU framework (Sec. 2), then discuss the adaption of a widely used algorithm for the Ising model simulation to better leverage TPU architecture (Sec. 3). We then follow up with benchmark results (Sec. 4) and in Section 5 highlight the implementation details and high-level performance analysis. Finally, we conclude the paper with Section 6, outlining our views on how the recent developments in both software and hardware for machine learning applications can impact the scientific computation applications.

\section{TPU Device Architecture}
TPU is a programmable linear algebra accelerator optimized for machine learning workloads. In the Cloud TPU v3 architecture, one TPU unit consists of four TPU chips on a board (we will use ``TPU unit'' or just ``unit'' to refer to such a board throughout this paper), and each TPU chip contains two TensorCores. Those TensorCores are treated as independent processors that communicate to each other through a dedicated high-bandwidth low-latency inter-chip network. In a larger system, more than 1000 TPU chips are packed on a two-dimensional toroidal mesh inter-chip interconnect network to form a TPU cluster known as a ``TPU Pod'' \citep{tpupod2019}.  In addition, each TPU unit is paired up with a TPU host server, with CPU, memory and disk resources attached. The communication links between TPU unit and its host server is through the regular data center connections. The host server can be used for data preparation, I/O task while TPU can be leveraged for compute-intensive tasks. The dedicated two-dimensional toroidal mesh allows all TensorCores in a ``Pod'' to work in lockstep efficiently, without going through the host servers. This large-scale high-bandwidth low-latency connectivity between TPU chips provides a significant advantage to achieve the strong linear-scaling performance reported in this paper.

\begin{figure*}[h!]
\centering
\includegraphics[width=0.8\textwidth]{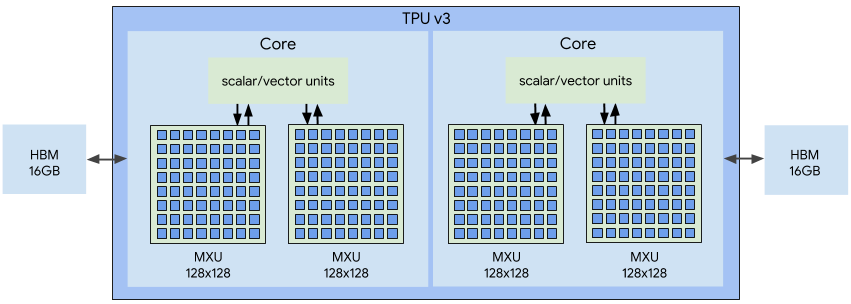}
\caption{One TPU chip consists of two TensorCores. A TensorCore in 3rd generation TPU (TPU v3) consists of a scalar processor, vector processor, and two matrix units. The arrows depict the datapaths across different processors/units and the high-bandwidth memory (HBM), the diagram is borrowed from \citep{tpusys2019}, here ``Core'' is the ``Tensorcore''.}
\label{fig:fish_chip}
\end{figure*}

The TensorCore, depicted in Figure~\ref{fig:fish_chip}, is optimized for dense linear algebra computations. It contains distinct classes of computing units, such as a scalar processor, a vector processor, accumulators and  matrix units \citep{tpusys2019, Jouppi2017, Jouppi2019}. All those processors are backed by its 16GB High-Bandwidth Memory (HBM). Vectorized operations are handled either in the vector processor directly or forwarded to corresponding extended vector units. Each extended vector unit takes the input operands, performs the corresponding operations, and returns the results back to the vector processor. One of these extended vector units is the matrix unit (MXU), which is capable of performing $128\times 128$ multiply-accumulate operations in each cycle \citep{tpuperf2019}. The MXU is the main computing power of the TPU architecture, so it should be exploited as much as possible. While its inputs and outputs are 32-bit floating point values, the MXU rounds inputs down to bfloat16---a 16-bit floating point representation that provides better training and model accuracy than the IEEE half-precision representation---before multiplying .

\begin{figure}[h!]
\centering
\includegraphics[scale=0.5]{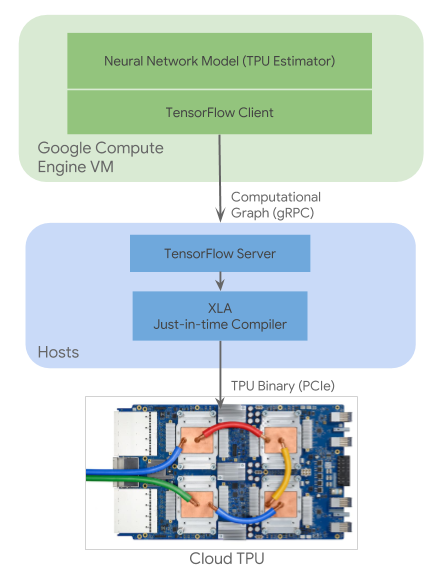}
\caption{TPU software architecture, consisting of the neural network model or other computational task, TensorFlow client, TensorFlow server and XLA compiler \citep{tpusys2019}.}
\label{fig:tensorcore}
\end{figure}

Machine learning research shows that many machine learning models can tolerate lower precision arithmetic without degradation of converged accuracy. TPU supports storing values in bfloat16 format as a way to reduce the size of data and allow larger models to fit in memory \citep{tpubfloat16}. For scientific computing, however, low precision is potentially dangerous because the increasing rounding error can introduce significant bias or divergent computation. In the case of the Ising model, while the binary spin values can be encoded in bfloat16 without loss of accuracy, the acceptance ratio and the random numbers used to determine acceptance in an MCMC simulation are more sensitive to reduced precision. However, our experiments show no noticeable differences in accuracy between bfloat16 and float32. By using bfloat16 instead of float32, we are able to simulate larger systems and leverage the MXU more effectively.

We program TPU through TensorFlow. The flow to run programs on TPU is roughly depicted in Figure~\ref{fig:tensorcore}. More details are available in TensorFlow's official XLA page\footnote{www.tensorFlow.org/xla}. In the first stage, TensorFlow constructs the computation graph and marks the graph for replication. Then the graph is rewritten to be TPU-compatible and compiled to a High Level Optimizer (HLO) program. Next, the Accelerated Linear Algebra (XLA) compiler takes over and converts HLO operations to Low Level Optimizer (LLO) code---effectively ``TPU assembly code'', which can be readily executed on TPU. The graph construction and the compilation occurs on host server and incurs over-head cost. But once the compiled LLO code is deployed to TPU's the computation step can repeat as many times as required without the intervention from the host servers.   

Importantly XLA also provides communication primitives such as CollectivePermute\footnote{https://www.tensorflow.org/xla/operation\_semantics\#collectivepermute} and AllToAll\footnote{https://www.tensorflow.org/xla/operation\_semantics\#alltoall}. These primitives are implemented over the dedicated high-bandwidth low-latency inter-chip interconnect network so the communication between any TPU chips within a TPU Pod is extremely efficient. 

Another important detail that can have a critical impact on performance and memory usage  is the choice of the shape of the tensor variables used in the program (expressed as TensorFlow tensors). According to the performance guide \citep{tpuperf2019}, unlike most other architectures, arrays in TPU are tiled in two dimensions. This entails to padding one dimension to a multiple of 8, and the other dimension to a multiple of 128. XLA performs data layout transformations and data are arranged in memory such that the hardware can efficiently process them. Programs that operate on array sizes undividable by 8 will have sub-optimal performance.

\section{The Ising Model}
Mathematically, the Ising spin Hamiltonian is given by
\[H(\pmb{\sigma}) = -J\sum_{\langle ij\rangle}\sigma_i\sigma_j - \mu\sum_{i=1}^N\sigma_i\]
where $\sigma_i$ is a random variable assuming the values of $\pm 1$ on sites $i = 1, \ldots, N$ of a $d$-dimensional hypercubic lattice, and $\langle ij\rangle$ indicates that sites $i$ and $j$ are nearest neighbors. (Note that throughout this paper, we will use the bold $\pmb{\sigma}$ to represent a tensor of spins, while the elements within it will be represented as regular $\sigma$). The first term, where the sum is over pairs of nearest-neighbor sites, represents the interaction energy that favors an ordered ferromagnetic state (if $J > 0$). The second term, involving the interaction between the applied field  and the spin system, is of a paramagnetic character. The configuration probability is given by the Boltzmann distribution with inverse temperature $\beta = (k_BT)^{-1}$:
\[\pi(\pmb{\sigma}) = \frac{e^{-\beta H(\pmb{\sigma})}}{Z_\beta}\]
where $Z_\beta = \sum_{\pmb{\sigma}}e^{-\beta H(\pmb{\sigma})}$ is the partition function and $k_B$ is the Boltzmann constant. For a function $f$ of the spins (``observable''), we denote $\langle f\rangle=\sum_{\pmb{\sigma}}f(\pmb{\sigma})\pi(\pmb{\sigma})$ the expectation (mean value) of $f$. $\langle f\rangle$ can often be difficult to evaluate numerically if there are many states in the system. The Markov Chain Monte Carlo is the most commonly used monte carlo algorithm to calculate statistics on the Ising model.

Without loss of generality, in what follows, we assume no external magnetic field, i.e., $\mu=0$ and $J=1$, and the 2D lattice has circular boundary or in other words, is a torus. A given configuration of the lattice (spin values) is represented by matrix $\pmb{\sigma}$.

\subsection{Checkerboard Algorithm}
The Metropolis-Hastings algorithm is the standard algorithm to simulate the Ising model. At each step, it proposes a candidate spin and flips the candidate based on the energy difference and acceptance probability. Closely related to this vanilla version that flips one spin at each step, there is another similar algorithm by flipping non-interacting spins in parallel, i.e., the efficient checkerboard algorithm \citep{Preis2009}. 
\begin{figure*}[h!]
\centering
\includegraphics[width=\textwidth]{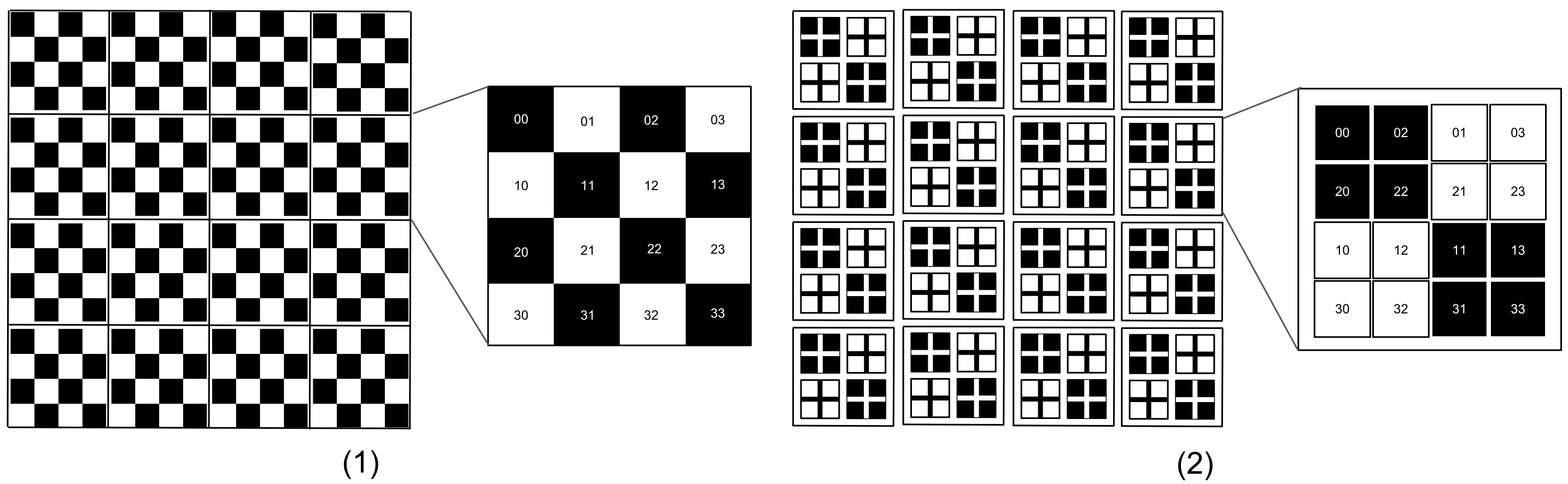}
\caption{A 2-d checkerboard: (1) Original checkerboard: on the left, the $16\times 16$ board is split into a $4\times 4$ grid of $4\times 4$ sub-lattices, i.e., it is represented by a $[4, 4, 4, 4]$ tensor, where $[l, k, :, :]$ is the sub-lattice at $[l, k]$ of the grid; on the right, the sub-lattice is zoomed in and the indices of its spin sites are shown; (2) Reorganized checkerboard: one the left, each $4\times 4$ sub-lattice is reorganized by 4 ``compact'' $2\times 2$ sub-lattices; on the right,  4 ``compact'' $2\times 2$ sub-lattices are zoomed in and their original indices from the $4\times 4$ sub-lattice are shown. In general, such alternate coloring of black and white can be extended to lattices with any dimensions.}
\label{fig:checkerboard}
\end{figure*}

Like the checkerboard above (Figure~\ref{fig:checkerboard}-(1)), spins in the lattice are colored black and white. The energy difference by flipping a spin of one color is completely described by its 4 neighbors of the opposite color. Thus, by fixing all spins of one color, the spins of the opposite color have no interactions with each other, and can be updated independently using Metropolis-Hastings. This observation leads to a highly parallel algorithm by alternating the 2 sub-routines below:
\begin{itemize}
\setlength\itemsep{0em}
\item Fixing all black spins, flip each white spin based on Metropolis Hastings in parallel.
\item Fixing all white spins, flip each black spin based on Metropolis Hastings in parallel. 
\end{itemize}
Write $\pi(\pmb{\sigma}) = \pi(\pmb{\sigma_w, \sigma_b})$, where $\pmb{\sigma_w}$ are values of all white spins and $\pmb{\sigma_b}$ are values of all black ones. The transition kernel is,
\begin{equation}
\begin{split}
&P\{(\pmb{\sigma_w, \sigma_b})\rightarrow(\pmb{\sigma_w^*, \sigma_b^*})\} \\
&= P\{(\pmb{\sigma_w, \sigma_b})\rightarrow(\pmb{\sigma_w^*, \sigma_b})\}P\{(\pmb{\sigma_w*, \sigma_b})\rightarrow(\pmb{\sigma_w^*, \sigma_b^*})\}
\end{split}
\end{equation}
and
\[P\{(\pmb{\sigma_w, \sigma_b})\rightarrow(\pmb{\sigma_w^*, \sigma_b})\}=\prod_{i\in\pmb{w}}\min\{1, e^{\beta(\sigma_i^* - \sigma_i)}\cdot \text{nn}(i)\}\]
\[P\{(\pmb{\sigma_w^*, \sigma_b})\rightarrow(\pmb{\sigma_w^*, \sigma_b^*})\}=\prod_{i\in\pmb{b}}\min\{1, e^{\beta(\sigma_i^* - \sigma_i)}\cdot \text{nn}(i)\}\]
where $\text{nn}(i)$ is the sum of neighbor spins of $i$. By conditional probability decomposition, it is easy to show that
\[\pi(\pmb{\sigma_w^*, \sigma_b^*}) = \sum_{\pmb{\sigma_w, \sigma_b}}P\{(\pmb{\sigma_w, \sigma_b})\rightarrow(\pmb{\sigma_w^*, \sigma_b^*})\}\pi(\pmb{\sigma_w, \sigma_b})\]
thus, the transition kernel satisfies the stationary distribution. The proof is based on Metropolis-Within-Gibbs Sampler \cite{Mueller1991} and is included in the supplemental materials. 

\subsection{Computation}
The compute-intensive part of the checkerboard algorithm is the computation of the sum of neighbor values for each spin. To leverage the MXU, for a given lattice with size $[128\times m, 128\times n]$ in a TPU core, we divide it into a $[m, n]$ grid of $128\times 128$ sub-lattices (Figure~\ref{fig:checkerboard}-(1)). Define the kernel matrix $K$ as,
\[K = \begin{bmatrix}
    0 & 1 & 0 & \dots  & 0 & 0 & 0 \\
    1 & 0 & 1 & \dots  & 0 & 0 & 0 \\
    \vdots & \vdots & \vdots & \ddots & \vdots & \vdots & \vdots \\
    0 & 0 & 0 & \dots & 1 & 0 & 1 \\
    0 & 0 & 0 & \dots & 0 & 1 & 0
\end{bmatrix}_{128\times 128}\]
then for a given sub-lattice $\pmb{\sigma}_{ij}$, $\text{matmul}(\pmb{\sigma}_{ij}, K) + \text{matmul}(K, \pmb{\sigma}_{ij})$ calculates the sum of nearest neighbors for all its internal sites. However, for the boundary sites, half of their nearest neighbors are missing from the sums. Those neighbors are on the boundaries of neighboring sub-lattices and ought to be sliced out and added into the sums. The whole lattice is updated one color at a time. To fix the spins of one color on the checkerboard, we multiply flip probabilities with a binary mask $M$, where
\[M = \begin{bmatrix}
    1 & 0 & 1 & \dots  & 0 & 1 & 0 \\
    0 & 1 & 0 & \dots  & 1 & 0 & 1 \\
    \vdots & \vdots & \vdots & \ddots & \vdots & \vdots & \vdots \\
    1 & 0 & 1 & \dots & 0 & 1 & 0 \\
    0 & 1 & 0 & \dots & 1 & 0 & 1
\end{bmatrix}_{128\times 128}\]
In summary, the algorithm to update the lattice given the color $c$ and configuration $\pmb{\sigma}$ is given in Algorithm \ref{algo:1}. By alternating colors with different $c$, the Ising model is properly simulated. However, in Algorithm \ref{algo:1}, there are several redundant calculations that are subject to further optimization: 1) Line 1 generates the probability for all spins, while only the spins colored by $c$ are eligible for flipping. 2) Lines 2-6 calculate the nearest neighbor sums of all spins, however, only spins colored by $c$ are updated. 3) Lines 8-9 generate flips but multiplying $mask$ to fix the opposite color is expensive. To eliminate all the redundancies above, we reorganize the lattice and represent it in a compact way, i.e., the lattice is instead split into a $[m', n']$ grid of $[256, 256]$ sub-lattices, and define 4 ``compact'' $[m', n', 128, 128]$ sub-lattices (Figure~\ref{fig:checkerboard}-(2)),
\[\pmb{\hat{\sigma}}_{00} = \pmb{\sigma}[:, :, 0::2, 0::2], \pmb{\hat{\sigma}}_{01} = \pmb{\sigma}[:, :, 0::2, 1::2]\]
\[\pmb{\hat{\sigma}}_{10} = \pmb{\sigma}[:, :, 1::2, 0::2], \pmb{\hat{\sigma}}_{11} = \pmb{\sigma}[:, :, 1::2, 1::2]\]
and kernel,
\[\hat{K} = \begin{bmatrix}
    1 & 1 & 0 & \dots  & 0 & 0 & 0 \\
    0 & 1 & 1 & \dots  & 0 & 0 & 0 \\
    \vdots & \vdots & \vdots & \ddots & \vdots & \vdots & \vdots \\
    0 & 0 & 0 & \dots & 0 & 1 & 1 \\
    0 & 0 & 0 & \dots & 0 & 0 & 1
\end{bmatrix}_{128\times 128}\]
thus, $\pmb{\hat{\sigma}}_{00}$ and $\pmb{\hat{\sigma}}_{11}$ are all ``black'' spins, and $\pmb{\hat{\sigma}}_{01}$ and $\pmb{\hat{\sigma}}_{10}$ are all ``white'' spins. Now, it is trivial to show that the nearest neighbor sums of internal spins of those 4 ``compact'' sub-lattices are
\begin{equation*}
\begin{split}
&\text{nn}(\pmb{\hat{\sigma}}_{00})=\text{matmul}(\pmb{\hat{\sigma}}_{01}, \hat{K}) + \text{matmul}(\hat{K}^T, \pmb{\hat{\sigma}}_{10}) \\
&\text{nn}(\pmb{\hat{\sigma}}_{11})=\text{matmul}(\hat{K}, \pmb{\hat{\sigma}}_{01}) + \text{matmul}(\pmb{\hat{\sigma}}_{10}, \hat{K}^T) \\
&\text{nn}(\pmb{\hat{\sigma}}_{01})=\text{matmul}(\pmb{\hat{\sigma}}_{00}, \hat{K}^T) + \text{matmul}(\hat{K}^T, \pmb{\hat{\sigma}}_{11}) \\
&\text{nn}(\pmb{\hat{\sigma}}_{10})=\text{matmul}(\hat{K}, \pmb{\hat{\sigma}}_{00}) + \text{matmul}(\pmb{\hat{\sigma}}_{11}, \hat{K})
\end{split}
\end{equation*}
and their boundary spins are corrected using a similar approach as in Algorithm \ref{algo:1}. The complete algorithm is presented in Algorithm \ref{algo:2}. According to our experiments, it is about 3x faster than Algorithm \ref{algo:1} and has less memory footprint as it uses less temporary HBM.

\begin{algorithm}
\SetAlgoLined
\SetKwInOut{Input}{input}
\SetKwInOut{Output}{output}
\Input{c: color ($black$ or $white$) \newline $\pmb{\sigma}\in \{0, 1\}^{m\times n\times 128\times 128}$: a configuration}
\Output{A new configuration}

\tcp{$\boldsymbol{\cdot}$ is element-wise multiplication, : or :$s$ is array slicing.}

$\text{probs} = \text{random\_uniform}([m, n, 128, 128])\in [0, 1)$

\tcp{$\text{nn}(\pmb{\sigma})$ is the sum of nearest neighbors of each site, it has the same size as the lattice. Here, $K$ is applied to each sub-lattice.}
$\text{nn}(\pmb{\sigma}) = \text{matmul}(\pmb{\sigma}, K) + \text{matmul}(K, \pmb{\sigma})$

\tcp{Compensate the northern boundaries of each sub-lattice.}
$\text{nn}(\pmb{\sigma})$[:, :, 0, :] += \{$\pmb{\sigma}$[-1:, :, -1, :], $\pmb{\sigma}$[:-1, :, -1, :]\}

\tcp{Compensate the southern boundaries of each sub-lattice.}
$\text{nn}(\pmb{\sigma})$[:, :, -1, :] += \{$\pmb{\sigma}$[1:, :, 0, :], $\pmb{\sigma}$[:1, :, 0, :]\}

\tcp{Compensate the western boundaries of each sub-lattice.}
$\text{nn}(\pmb{\sigma})$[:, :, :, 0] += \{$\pmb{\sigma}$[:, -1:, :, -1], $\pmb{\sigma}$[:, :-1, :, -1]\}

\tcp{Compensate the eastern boundaries of each sub-lattice.}
$\text{nn}(\pmb{\sigma})$[:, :, :, -1] += \{$\pmb{\sigma}$[:, 1:, :, 0], $\pmb{\sigma}$[:, :1, :, 0]\}

$\text{acceptance\_ratio} = \exp(-2\cdot \text{nn}(\pmb{\sigma}) \cdot \pmb{\sigma})$

$\text{mask} = M$ if $c$ is $black$ else $1 - M$

$\text{flips} = (\text{probs} < \text{acceptance\_ratio}) \cdot \text{mask}$

\Return $\pmb{\sigma} - 2\cdot \text{flips}\cdot \pmb{\sigma}$
\caption{Subroutine \textbf{UpdateNaive}($c$, $\pmb{\sigma}$)}
\label{algo:1}
\end{algorithm}

\begin{algorithm}
\SetAlgoLined
\SetKwInOut{Input}{input}
\SetKwInOut{Output}{output}
\Input{c: color ($black$ or $white$) \newline $\pmb{\hat{\sigma}}_{00}, \pmb{\hat{\sigma}}_{01}, \pmb{\hat{\sigma}}_{10}, \pmb{\hat{\sigma}}_{11}\in \{0, 1\}^{m'\times n'\times 128\times 128}$: a configuration}
\Output{A new configuration}

$\text{probs0} = \text{random\_uniform}([m', n', 128, 128])\in [0, 1)$

$\text{probs1} = \text{random\_uniform}([m', n', 128, 128])\in [0, 1)$

\eIf{black}{
$\pmb{\hat{\sigma}}_{0} = \pmb{\hat{\sigma}}_{00}$

$\pmb{\hat{\sigma}}_{1} = \pmb{\hat{\sigma}}_{11}$

$\text{nn0}=\text{matmul}(\pmb{\hat{\sigma}}_{01}, \hat{K}) + \text{matmul}(\hat{K}^T, \pmb{\hat{\sigma}}_{10})$

$\text{nn0}$[:, :, 0, :] += \{$\pmb{\hat{\sigma}}_{10}$[-1:, :, -1, :], $\pmb{\hat{\sigma}}_{10}$[:-1, :, -1, :]\}

$\text{nn0}$[:, :, :, 0] += \{$\pmb{\hat{\sigma}}_{01}$[:, -1:, :, -1], $\pmb{\hat{\sigma}}_{01}$[:, :-1, :, -1]\}

$\text{nn1}=\text{matmul}(\hat{K}, \pmb{\hat{\sigma}}_{01}) + \text{matmul}(\pmb{\hat{\sigma}}_{10}, \hat{K}^T)$

$\text{nn1}$[:, :, -1, :] += \{$\pmb{\hat{\sigma}}_{01}$[1:, :, 0, :], $\pmb{\hat{\sigma}}_{10}$[:1, :, 0, :]\}

$\text{nn1}$[:, :, :, -1] += \{$\pmb{\hat{\sigma}}_{01}$[:, 1:, :, 0], $\pmb{\hat{\sigma}}_{01}$[:, :1, :, 0]\}
}{
$\pmb{\hat{\sigma}}_{0} = \pmb{\hat{\sigma}}_{01}$

$\pmb{\hat{\sigma}}_{1} = \pmb{\hat{\sigma}}_{10}$

$\text{nn0}=\text{matmul}(\pmb{\hat{\sigma}}_{00}, \hat{K}^T) + \text{matmul}(\hat{K}^T, \pmb{\hat{\sigma}}_{11})$

$\text{nn0}$[:, :, 0, :] = \{$\pmb{\hat{\sigma}}_{11}$[-1:, :, -1, :], $\pmb{\hat{\sigma}}_{11}$[:-1, :, -1, :]\}

$\text{nn0}$[:, :, :, -1] = \{$\pmb{\hat{\sigma}}_{00}$[:, 1:, :, 0], $\pmb{\hat{\sigma}}_{00}$[:, :1, :, 0]\}

$\text{nn1}=\text{matmul}(\hat{K}, \pmb{\hat{\sigma}}_{00}) + \text{matmul}(\pmb{\hat{\sigma}}_{11}, \hat{K})$

$\text{nn1}$[:, :, -1, :]=\{$\pmb{\hat{\sigma}}_{00}$[1:, :, 0, :], $\pmb{\hat{\sigma}}_{00}$[:1, :, 0, :]\}

$\text{nn1}$[:, :, :, 0]=\{$\pmb{\hat{\sigma}}_{11}$[:, -1:, :, -1], $\pmb{\hat{\sigma}}_{11}$[:, :-1, :, -1]\}

}

$\text{acceptance\_ratio0} = \exp(-2\cdot \text{nn0} \cdot \pmb{\hat{\sigma}}_{0})$

$\text{acceptance\_ratio1} = \exp(-2\cdot \text{nn1} \cdot \pmb{\hat{\sigma}}_{1})$

$\text{flips0} = (\text{probs0} < \text{acceptance\_ratio0})$

$\text{flips1} = (\text{probs1} < \text{acceptance\_ratio1})$

\Return ($\pmb{\hat{\sigma}}_{0} - 2\cdot \text{flips0}\cdot \pmb{\hat{\sigma}}_{0}$),  ($\pmb{\hat{\sigma}}_{1} - 2\cdot \text{flips1}\cdot \pmb{\hat{\sigma}}_{1}$)
\caption{Subroutine \textbf{UpdateOptim}($c$, $\pmb{\sigma}$)}
\label{algo:2}
\end{algorithm}


\section{Simulation Results}
\subsection{Correctness}
The average magnetization per spin for a given $\beta$ is defined as,
\[m(T) = m(\beta) =\langle\pmb{\sigma}\rangle=\frac{1}{N}\sum_i\sigma_i\]
where $N = n^2$ for a square lattice with size $n$. It can be shown that below the critical temperature, i.e.,
$(k_B\beta)^{-1} = T < T_c = (k_B\beta_c)^{-1}=\frac{2}{k_B\ln(1 + \sqrt{2})}$, there is spontaneous magnetization---the interaction among spins is sufficiently large to cause neighboring spins to spontaneously align. On the other hand, thermal fluctuations completely eliminate any alignment above the critical temperature. Moreover, at the critical temperature, there is a discontinuity in the first derivative of $\langle m(T)\rangle$ with respect to $T_c$. This discontinuity generates a downward drop in the average magnetization. The sudden loss of spontaneous magnetization above the critical temperature is a signature of phase transition. Besides average magnetization, a more sensitive test of correctness is the Binder parameter \citep{Binder1993}, which is given by
\[U_4(T) = 1 - \frac{\langle m(T)^4\rangle}{3\langle m(T)^2\rangle^2}\]
namely, the \emph{kurtosis} of $m(T)$. It is frequently used to accurately determine phase transition points in numerical simulations of various models.

\begin{figure*}[h!]
\centering
\includegraphics[width=\textwidth]{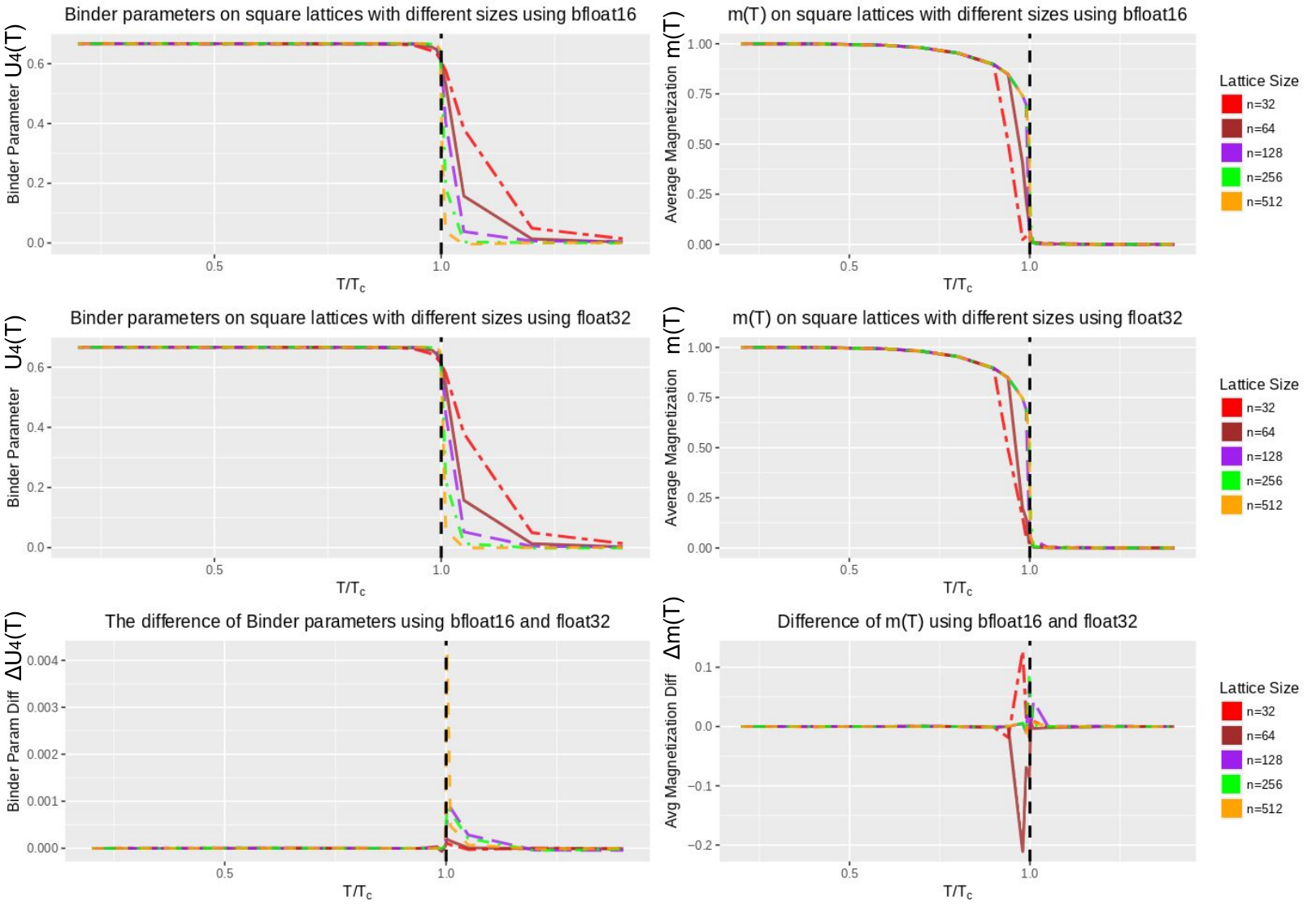}
\caption{Binder parameter $U_4(T)$ and average magnetization $m(T)$ as a function of $T/T_c$ for various sizes of the two dimensional square lattice Ising model. Each data point is calculated by a Markov Chain of 1,000,000 samples using checkerboard update, where the first 100,000 samples are discarded for burn-in, and the rest 900,000 samples are used for the calculation. The plots of $U_4(T)$ for various lattice sizes cross almost perfectly at the critical temperature, which is shown additionally as a black dashed line, and their float32 and bfloat16 versions almost completely match. The plots of $m(T)$ show vanishing magnetization above critical temperature, but there are subtle differences between float32 and bfloat16 as $m(T)$ is a less sensitive test.}
\label{fig:correctness}
\end{figure*}

We verify the correctness of our algorithm and implementation by computing both quantities at various temperatures on different sizes of square lattices. As shown in Figure~\ref{fig:correctness}, the curves of average magnetization, with subtle differences, overlap with each other, and those of Binder parameters cross the critical line almost perfectly. Additionally, we investigate the implications of lower precision, i.e., bfloat16, for the estimation accuracy. In MCMC simulation of Ising model, lower precision has impact on the calculation of acceptance ratio and the random number generation, the bias introduced might accumulate and decrease the overall accuracy. However, in our experiments, all curves generated in bfloat16 and float32, especially those of Binder parameters, are almost the same and sharp turns around critical line are clearly observable. Based on this evidence, we argue that using bfloat16 has negligible impact on Ising model simulation, and in turn it offers two benefits: 1). We are able to simulate a larger lattice on a single TPU core because of smaller memory footprint, and 2). bfloat16 matrix multiplication with 32-bit accumulation is very efficient in MXU, while float32 matrix multiplication is more expensive as several bfloat16 passes are required.

\subsection{Benchmarks}
Preis et al. \citep{Preis2009} showed impressive acceleration in their single GPU implementation of checkerboard algorithm using the compute unified device architecture (CUDA). By exploiting GPU's  large pool of threads, its single-instruction multiple thread (SIMT) unit and memory hierarchy, their algorithm achieved 60x speedup on single GPU compared to its CPU counterpart. In their follow-up paper \citep{Block2010}, their original algorithm was modified to overcome the memory limitations of single GPU. The improved algorithm was able to simulate significantly larger systems and reached a performance of $\sim7.9774$ flips/ns throughput in its best performing variant. By combining CUDA with MPI on the CPU level, their distributed algorithm achieved 206 flips/ns on a $800,000^2$ lattice. Besides the work on GPU, another encouraging line of research is the use of field-programmable gate array (FPGA). A recent implementation achieves $\sim 614.4$ flips/ns throughput, see \citep{Ortega-Zamorano2016} and references therein.

To quantify the performance of our implementation, we run our algorithm on TPU v3 using single core and multiple cores on TPU v3 clusters. As in \citep{Preis2009, Block2010, Ortega-Zamorano2016}, we measure the time spent on one sweep update, i.e., one update on all ``black'' spins plus one update on all ``white'' ones, and compute the average number of flips per nanosecond by dividing with $n^2$.
\subsubsection{Single TPU Core}
First, we simulate the system on a single TPU v3 core (half TPU v3 chip). Our choice of lattice size is compatible with MXU's registers and achieves 100\% memory capacity utilization according to our profilings. We can simulate lattice  with size up to $(656\times 128)^2 = 83,968^2$, which consumes 96\% of the memory. The benchmarks in various sizes from $(20\times 128)^2$ to $(640\times 128)^2$ are summarized in Table.~\ref{table:1core}, the lattice size and flips in nanoseconds increase in tandem, as more computation is spent on matrix multiplication. For comparison, we also implemented the algorithms based on \citep{Preis2009, Block2010} under \texttt{CUDA 10.1}, and using its \texttt{cuRand} package for random number generation and its \texttt{Thrust} package for reductions. To avoid the excessive temporary memory allocation on GPU, we wrote a custom memory allocator to reuse temporary memory. The benchmark we obtained on NVIDIA's Tesla V100, which powered by NVIDIA's latest Volta architecture, is 11.3704 flips/ns. Other older benchmarks published in \citep{Preis2009, Block2010, Ortega-Zamorano2016} are also listed for reference. 

\begin{table*}[ht]
\begin{minipage}[c]{0.47\linewidth}\centering
\begin{tabular}{|c|c|c|}
  \hline
  lattice size $n^2$ &  
  (flips/ns)  &  (nJ/flip)  \\
  \hline
  \hline
  $(20\times 128)^2$ &  8.1920 & 12.2070 \\
  $(40\times 128)^2$ & 9.3623 & 10.6811 \\ 
  $(80\times 128)^2$& 12.3362 & 8.1062\\
  $(160\times 128)^2$& 12.8266 & 7.7963\\
  $(320\times 128)^2$& 12.9056 & 7.7486\\
  $(640\times 128)^2$& 12.8783 & 7.7650 \\
  \hline
  \hline
  GPU in \citep{Preis2009, Block2010} &  \textbf{7.9774} & -- \\
  Nvidia Tesla V100 & \textbf{11.3704} & 21.9869 \\
  FPGA in \citep{Ortega-Zamorano2016} &  \textbf{614.4} & -- \\
  \hline
\end{tabular}
\end{minipage}
\begin{minipage}[c]{0.49\linewidth}
\centering
\includegraphics[width=1\linewidth]{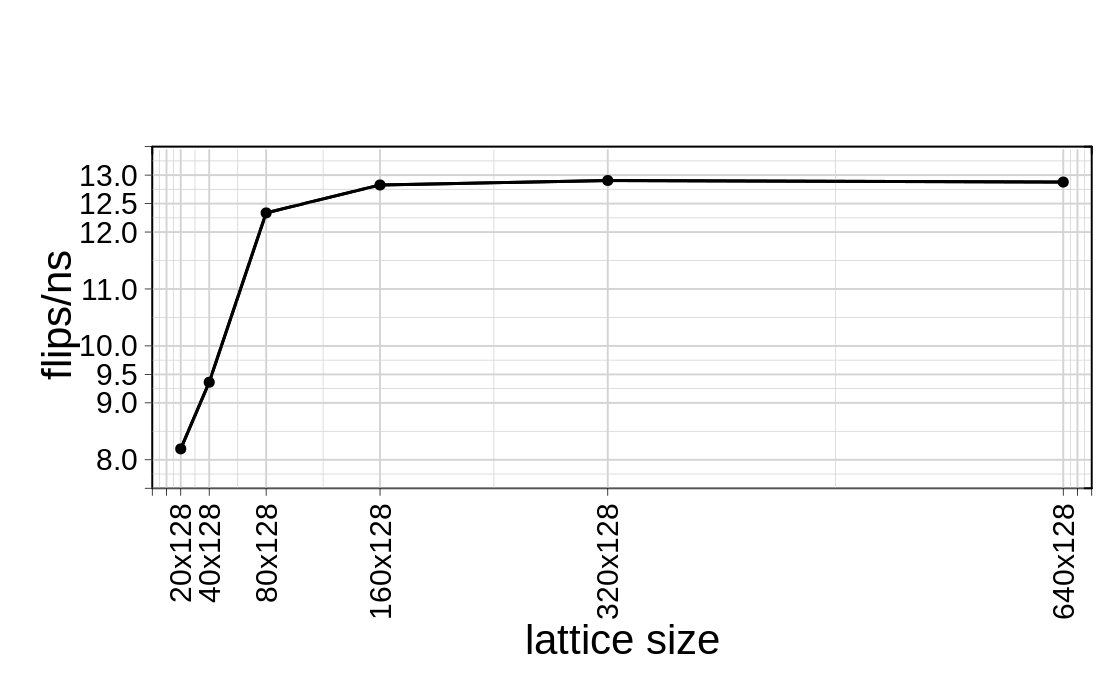}
\end{minipage}
\caption{The computation throughput (flips/ns) and the estimated energy consumption upper bound (nJ/flip) with different sizes of the square lattice on a single TPU v3 core (half TPU v3 chip). Not comparing to FPGA, a single TensorCore sustains more flips/ns at all but the two smallest lattice sizes and consistently shows better energy efficiency.}
\label{table:1core}
\end{table*}

It is also interesting to estimate the energy consumption in the computation. Assuming the average power consumption during the operation to be $\boldsymbol{\mathcal{P}}$W and the throughput achieved to be $\boldsymbol{\mathcal{F}}$flips/ns, the corresponding energy used is $(\boldsymbol{\mathcal{P}}/\boldsymbol{\mathcal{F}})$nJ/flip. The actual average power consumption depends on many factors and usually sophisticated modeling and measurements are needed.
However, for a rough estimate of the upper bound, we use $250$ W for GPU (based on NVIDIA's Tesla V100 Spec for PCIe version max power consumption\footnote{\raggedright\url{images.nvidia.com/content/technologies/volta/pdf/437317-Volta-V100-DS-NV-US-WEB.pdf}}). 
While Google doesn't publish the number for TPU v3, it has been estimated to be $200$W\footnote{\raggedright\url{www.nextplatform.com/2018/05/10/tearing-apart-googles-tpu-3-0-ai-coprocessor/}} for a TPU v3 chip, or equivalently $100W$ for a TPU v3 core.

\subsubsection{Linear Scaling on Multiple TPU cores}
The idea to simulate Ising model on TPU v3 clusters is to split the whole lattice into sub-lattices, and exchange their boundary values to calculate the nearest-neighbor sums and update the sub-lattices in each core independently. The challenges are to handle the synchronization among the cores: 1). Block the update of sub-lattices until all the nearest-neighbor sums are calculated. 2). Block next iteration before the update of all sub-lattices is completed. Fortunately, such synchronization is already implemented in TensorFlow op \texttt{collective\_permute}, which is used to exchange data by specifying source and target cores. 

In its current release, TPUs in a TPU Pod are organized into a grid and each TPU core has an associated coordinate. All cores communicate through a specialized high-speed interconnect. In our experiments, we use smaller sections of a pod called \emph{slices} \citep{tpupod2019}, and show that the TPU Pod interconnect makes the overhead of exchanging boundary values between cores negligible. As a result, we observe strict linear scaling of flips/ns to the number of TPU cores, as shown in Table.~\ref{table:ncore}.

In \citep{Block2010}, the communications between GPUs are handled by MPI through the hosts, which is potentially a bottleneck in the simulations. Another interesting comparison is to benchmark their multi-GPU algorithm using \texttt{Nvidia NVLink Fabric}, which enables the interconnect of 8 GPUs, or \texttt{Nvidia NVSWITCH} that support the interconnect of 16 GPUs \citep{nvlink2019}. However, as the code for multi-GPU simulation is not available and the limitations on the number of interconnect GPUs, we leave it for our future work.

\begin{table*}
\centering
\begin{tabular}{|c|c|c|c|c|}
  \hline
  \#cores & lattice size & \begin{tabular}{c} time of whole \\ lattice update (ms) \end{tabular} & \begin{tabular}{c}throughput \\(flips/ns) \end{tabular}& \begin{tabular}{c}energy consumption \\ (nJ/flip)\end{tabular}\\
  \hline
  \hline
  $1\times 1\times 2$ & $(896\times 128)^2$ & 574.7 & 22.8873 & 8.7385\\
  $2\times 2\times 2$ & $(1792\times 128)^2$ & 574.9 & 91.5174 & 8.7415\\
  $4\times 4\times 2$ & $(3584\times 128)^2$ & 575.0 & 366.0059 & 8.7430\\
  $8\times 8\times 2$ & $(7168\times 128)^2$ & 575.2 & 1463.5146 & 8.7461\\
  $16\times 16\times 2$ & $(14336\times 128)^2$ & 575.3 & 5853.0408 & 8.7476 \\
  \hline
  \hline
  64 GPUs \citep{Block2010} & $800,000^2$ & $\sim 3000$ & \textbf{206} & --\\
  \hline
\end{tabular}
\caption{Each core contains a $[896 \times 128, 448 \times 128]$ sub-lattice. Hence, for a $n\times n \times 2$-core cluster, the lattice size is $(512 \times 128\times n)^2$. Dividing flips/ns by number of cores, the flips per nanosecond per TPU v3 core is roughly 11.4337, compared to 3.2188 per GPU---a 250\% speedup. Note that the energy consumption estimate is an upper bound.}
\label{table:ncore}
\end{table*}

\section{Implementation Highlights and Performance Analysis}
In the previous section, we report the competitive performance numbers achieved in our experiments. It is worthwhile going into some depth to highlight the implementation design choices and analyze the performance, in order to provide more insights on this new approach towards scientific simulations. The entire source code is also available through AD appendix.    
\subsection{Implementation Highlights: TensorFlow, SIMD, Highspeed Mesh Network}
While most commonly seen applications of TensorFlow are in the context of machine learning, the functionalities available are highly relevant to various scientific applications. At a high level, TensorFlow provides various transformation operations that can be applied on tensors. These tensors can be either stateful variables or temporary values. 

It is easy to see that the two-dimensional spin sites can be easily represented as a stateful tensor variable. In this work, we have chosen to use arrays of rank-4 tensors to represent the supergrid structure used in checkerboard algorithm ~\ref{algo:2}. More specifically, the partial spin lattice on each core is represented as a four-dimensional array: $\text{super\_grids}[N_x, N_y, 2, 2]$ where each of the array elements is a rank-$4$ tensor variable with shape of $[m, n, 128 \times i, 128 \times j]$. 

The tensor shape is chosen so the last two dimensions are always integer multiples of $128$, to better match TPU HBM tiling and the MXU structure\citep{tpuperf2019}. We also choose to use $\text{super\_grids}[:, :, 0, 0]$, $\text{super\_grids}[:, :, 1, 1]$ to represent the black compact sub-lattices, and use $\text{super\_grids}[:, :, 0, 1]$ and $\text{super\_grids}[:, :, 1, 0]$ to represent the white compact sub-lattices, as depicted in Figure~\ref{fig:checkerboard}-(2). In the results reported in Table~\ref{table:ncore} we use $(N_x, N_y, m, n, i, j) = (2, 2, 224, 112, 1, 1)$, which gives us the per-core lattice size of $128\times[896, 448]$.

To determine the acceptance for the flipping of each spin site, a  random tensor generation operation available in TensorFlow is used: $\text{tf.random\_uniform}$\footnote{\raggedright\url{www.tensorflow.org/api_docs/python/tf/random/uniform}}.
It generates random tensor for a given shape with uniform probability between $[0, 1]$ for each element. While this process is not the most compute-intensive, it does take up about $\sim10\%$ of the step time (more discussions on this in next sub-section).

The most compute-intensive part of the simulation is the computation of acceptance ratio, which involves summing on the nearest neighbor spin values. As pointed out previously, we leverage MXU's matrix multiplication to achieve this. Since each TensorCore provides a raw computational power of $\sim50$ TFLOPS, it greatly helps the efficiency of our simulation. This part takes up about $\sim60\%$ of the step time.

In TensorFlow framework, the expression of the computation is converted into a graph, and for TPU, it is further compiled through XLA into the executable programs (LLO) and deployed to TPU during run time. This Just-In-Time (JIT) compilation can incur overhead but usually it is fairly small (under a few secs) for smaller problems, and while it can sometimes take longer time (up to minutes) for larger problems, usually it is well-amortized for these larger problems as typically millions of steps are executed.

For the distributed multi-core case, TensorFlow also provides the primitive for SIMD programming on TPU. Using $\text{tf.tpu.replicate}$\footnote{\raggedright\url{www.tensorflow.org/api_docs/python/tf/tpu/replicate}}, one can easily replicate the computation across multiple TPU TensorCores with simple syntax (typically a few lines of code) to map and assign core ids using an object returned during the TPU system initialization call, which encapsulates the mesh topology information.  

\begin{figure}[h!]
\centering
\includegraphics[width=0.65\textwidth]{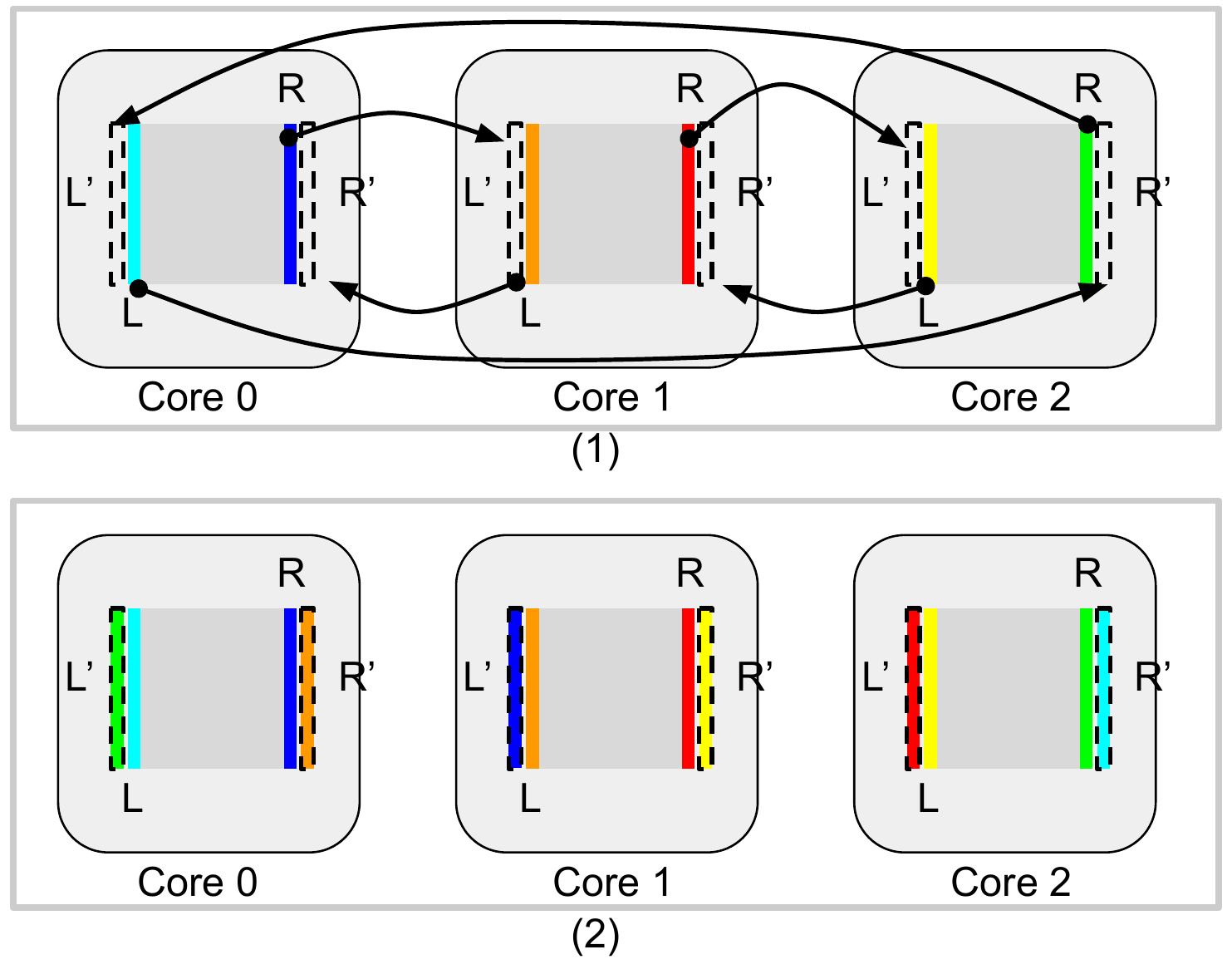}
\caption[]{Illustration of tensor values permute across 3 TensorCores and the procedure of acquiring the neighboring boundary grid values for each TensorCore. Each TensorCore contains a local sub-grid. The right boundary and left boundary are represented by tensors $\text{R}$ and $\text{L}$. And the extended boundaries that would need to be filled with the values from their neighbors are represented by tensors $\text{R}^\prime$ and $\text{L}^\prime$. (1) Showing the tensor values permutation with \\
$L^\prime=\mathrm{tpu\_ops.collective\_permute}(R, [[0, 1], [1, 2], [2, 0]])$ 
and \\ $R^\prime=\mathrm{tpu\_ops.collective\_permute}(L, [[0, 2], [2, 1], [1, 0]])$.
\\ (2) After the permutation, each core gets the boundary values from the neighboring core and the extended boundaries are filled in with the correct values on each core.}
\label{fig:collective_permute}
\end{figure}

Another critical component for the distributed case is that in the acceptance ratio computation, each core needs to exchange its border spin values with the neighboring cores. Again, TensorFlow provides a simple primitive that is syntactically simple and leverages TPU Pod's high speed dedicated inter-chip mesh network, $\text{tpu\_ops.collective\_permute}$\footnote{\raggedright\url{www.github.com/tensorflow/tensorflow/blob/master/tensorflow/python/tpu/ops/tpu_ops.py}} which allows a tensor's value to be permuted through different cores according to the source-destination pair specification. Note that in this case, each core has the same instruction and in the call to $\text{tpu\_ops.collective\_permute}$, the source-destination mapping contains globally identical specifications. Each core that is involved in the operation, when executing this part of program, will block until it sends and receives the corresponding values according to the source-destination specification. 

Figure~\ref{fig:collective_permute} shows an example of a collection of $3$ cores that are exchanging the 'boundaries' with periodic boundary condition. Each core has tensors $\text{R}$ and $\text{L}$ representing the right and left boundaries of the internal grid. Figure~\ref{fig:collective_permute} shows how each core can acquire the extended boundaries from its neighboring cores. The highly efficient inter-core communications in TPU Pod allows all TensorCores to work in lockstep with minimal latency even when a large number of cores participate in the communications. In fact, the time spent on this step in our experiments is well below $0.15\%$ in all cases and this explains the linear-scaling performance we observed. Note that in all our experiments, no attempt is made to match the logical layout of the lattice with the physical TPU cores: two logically neighboring sub-lattices could be distributed to two cores physically far apart, requiring multiple hops for the communication between them and, yet this doesn't cause any noticeable performance loss.      

\subsection{Performance Analysis}
\begin{figure}[h!]
\centering
\includegraphics[width=0.65\textwidth]{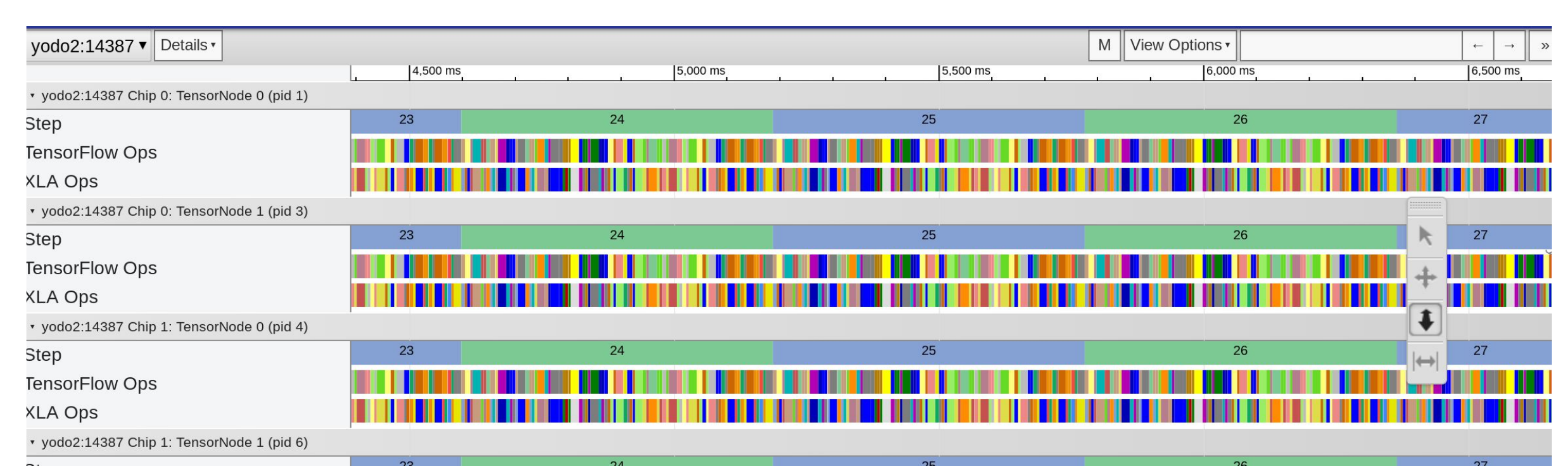}
\caption[]{A screen grab of the TPU profiling tool's trace viewer. Shown here is for the case with $16 \times 16 \times 2$ cores and only showing the traces from a few cores. It can also be seen that these cores progress in a lockstep fashion.}
\label{fig:profiler_trace}
\end{figure}
We look into several key components of the computations and their respective performance in terms of the time spent on each. We use the tool devloped for TPU profiling (available also in Google Cloud\footnote{\raggedright\url{cloud.google.com/tpu/docs/cloud-tpu-tools}}. The profiling tool is able to provide fine-grained analysis of the utilization of the hardware, the efficiency of the operations at program level and more. Figure~\ref{fig:profiler_trace} gives an example output from the tool's trace viewer. 
\begin{table*}
\centering
\begin{tabular}{|c|c|c|c|c|c|}
\hline
  \#cores & lattice size & \begin{tabular}{c} MXU \\ time (\%) \end{tabular}& \begin{tabular}{c}VPU \\ time (\%) \end{tabular}& \begin{tabular}{c}data \\ formatting \\ time (\%) \end{tabular} & \begin{tabular}{c}  collective\_permute \\ time (\%) \end{tabular} \\
  \hline
  $1\times 1\times 2$ & $(896\times 128)^2$ & 59.6 & 12 & 28.2 & 0.024 \\
  $2\times 2\times 2$ & $(1792\times 128)^2$ & 59.6 & 12 & 28.1 & 0.038\\
  $4\times 4\times 2$ & $(3584\times 128)^2$ & 59.5 & 11.9 & 28.2 & 0.063\\
  $8\times 8\times 2$ & $(7168\times 128)^2$ & 59.5 & 12 & 28.1 & 0.08\\
  $16\times 16\times 2$ & $(14336\times 128)^2$ & 59.4 & 12 & 28.1 & 0.11 \\
  \hline
\end{tabular} 
\caption{Percentage time breakdown of the computation. Note that since the step time for all cases are basically all $\sim580$ ms, the numbers in the table are proportional to absolute time. MXU time is most from the matrix multiplication operations employed for nearest-neighbor computation and it is the biggest portion. VPU time is the time spent in vector unit. In this case, it is mostly for generating random uniform tensors. Data formatting is the time spent on moving data, reshaping tensors, slicing etc. within a core. The inter-core communication time is very negligible for all cases and this is consistent with the strong linear-scaling we observed. In all cases, the amount of the data (the edges) that are moved between cores are $896\times128\times2=229,376$ bytes per edge in one direction and $448\times128\times2=114,688$ bytes per edge in another direction, for each core. These are very small data nd the observed time are primarily dominated by other factors such as synchronization and latency.}
\label{table:time_breakdown}
\end{table*}

Using the tool, we took measurements of the breakdown of the key operations at the HLO level: the time spent on the computation of the nearest neighbor sum (mostly using MXU), the time spent on generating the random uniform tensors (mostly using VPU), the time spent of data formatting and reshaping for the computation, and the time spent on data exchange between cores. Detailed breakdown is in Table~\ref{table:time_breakdown}. Note that from Table~\ref{table:ncore} we know that for all cases, the time step is essentially identical ($\sim580$ms), so percentage numbers in this table can be used directly for comparison across different cases. 
\begin{table}
\centering
\begin{tabular}{|c|}
\hline
(step time, collective\_permute time) \\ with various per-core lattice size (ms) \\
\begin{tabular}{|c|c|c|c|}
\hline
  \#cores & \begin{tabular}{c} $[896\times128$, \\ $ 448\times128]$ \end{tabular} & 
  \begin{tabular}{c} $[448\times128$, \\ $224\times128]$ \end{tabular} & 
  \begin{tabular}{c}
  $[224\times128$, \\ $112\times128]$ \end{tabular} \\ 
  \hline
  $4\times 4\times 2$ & (575.0, 0.37) & (255, 0.36) & (64.61, 0.18) \\
  $8\times 8\times 2$ & (575.2, 0.47) & (255.11, 0.41) & (64.69, 0.25) \\
  $16\times 16\times 2$ & (575.3, 0.65) & (255.03, 0.64) & (64.92, 0.58) \\
  \hline
\end{tabular} \\
\hline
\end{tabular} \\
\caption{The measured step time and collective\_permute time (in ms) at various per-core lattice size. The data amount exchanged are small as only edges of the sub-lattice are exchanged between cores so the measured time is not bandwidth limited (the largest edge is only $229,376$ bytes, and would take just $\sim0.023$msec to transport on a relatively moderate $10$GB network). We also see that for a given total lattice size, as the number of cores increase, the step time decreases.}
\label{table:comm2}
\end{table}

The first observation is that the time breakdown is fairly stable across different scales. We also see that the most expensive part is the computation of nearest neighbor sum, leveraging the MXU, which accounts for $\sim60\%$ of the time. The time spent on generating random tensors is also significant: it is  $\sim12\%$ of the time. The data formatting takes up $\sim28\%$ of the time. It is also worth noting that this part can be significantly worse if the shape of the tensor variables do not conform to the tiling in TPU HBM. The most interesting part is the time spent on inter-core communication, and it can be seen that it is taking very insignificant amount of time, even for the very large case when $512$ cores are involved. It is a key property of TPU Pod that allows the linear scaling. Note that since the cores participating in the ``collective\_permute'' operations will perform both sending and receiving and the time measured includes both the synchronization overhead between the cores as well as the time that the data travels between cores. However, since the data amount (the edges of the sub-lattice) is small, the time is not dominated by the data propagation and not bandwidth bound (the largest edge is only $229,376$ bytes, and would take just $\sim0.023$msec to transport even on a relatively moderate $10$GB network). Table ~\ref{table:comm2} shows the step time and the ``collective\_permute'' time with various per-core lattice size at different number of cores. The ``collective\_permute'' time in all cases are insignificant when compared with the step time. We also note that this time is more directly affected by the number of cores than the size of the sub-lattice, indicating that the communication is not in the bandwidth limited regime. 

Another interesting aspect of the data in Table~\ref{table:comm2} is that, for a constant full lattice size (i.e., the entries along the diagonal direction in Table~\ref{table:comm2}), while the step time consistently decreases as the number of cores increases, we can see that there are two different regimes of the rate of step time decrease. When the per-core lattice size decreases from $[896\times128, 448\times128]$ to $[448\times128, 224\times128]$, a $4\times$ decrease, the step time decreases from $\sim575$ ms to $\sim255$ms, or $\sim44\%$, instead of $25\%$. But when the per-core lattice size decrease from $[896\times128, 448\times128]$ to $[224\times128, 112\times128]$, the step time changes from $\sim255$ ms to $\sim65$ms, or $\sim25.5\%$. This is due to higher MXU utilization when the per-core lattice size is $[896\times128, 448\times128]$, and the utilization pattern changes (lower MXU utilization) when the per-core lattice size changes to $[448\times128, 224\times128]$, and stays about the same when the per-core lattice size is $[224\times128, 112\times128]$. 

\begin{table}
\begin{minipage}[c]{1.0\linewidth}\centering
\begin{tabular}{|c|c|c|}
  \hline
  \#cores & \% of roofline optimal & \% of HW peak  \\
  \hline
  \hline
  $1\times 1\times 2$ & 76.68 & 9.31\\
  $2\times 2\times 2$ & 76.65 & 9.3\\
  $4\times 4\times 2$ & 76.51 & 9.28\\
  $8\times 8\times 2$ & 76.52 & 9.27\\
  $16\times 16\times 2$ & 76.43 & 9.26\\
  \hline
\end{tabular}
\end{minipage}
\begin{minipage}[c]{0.9\linewidth}
\centering
\includegraphics[width=0.9\linewidth]{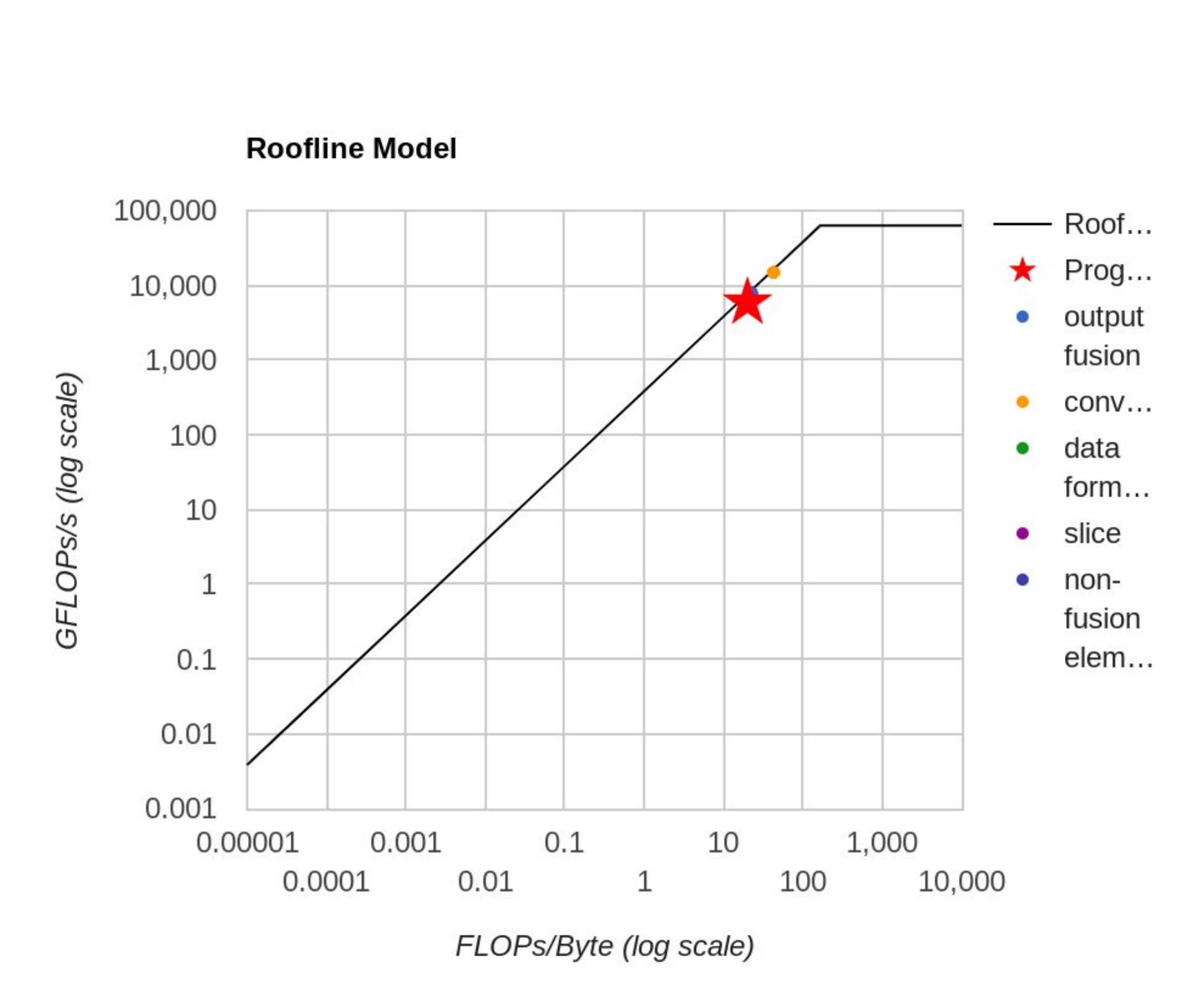}
\end{minipage}
\caption{Top: Achieved program FLOPS compared against the roofline model optimal performance and hardware peak FLOPS. All measured with per-core lattice size of $128 \times [896, 448]$. All measurements shown here are memory bound. Bottom: the roofline model plot for $16\times 16\times 2$ cores with per-core lattice size of $128 \times [896, 448]$.}
\label{table:flops_achieved}
\end{table}

Finally, we use the profiling tool to measure the FLOPS performance of our program. The results are summarized in Table~\ref{table:flops_achieved}. In all cases, the achieved FLOPS is about $76\%$ of the roofline (memory bound), or roughly $5.89$ TFLOPS. A rough estimate can be done using the number operations in matrix multiplication per core used for nearest neighbor sum: there are total $4\times896\times448$ of matrix multiplications of size $128 \time 128$ ($2\times$ from the inner grid nearest neighbor computations, and $2\times$ from the boundary nearest neighbor). Total number of operations is $896\times448\times128^3$. Using step time $\sim580$ms, we got the estimate of $5.8$ TFLOPS, vey close to the measured program FLOPS. It is also interesting to note that, from the slope of the roofline plot in Table~\ref{table:flops_achieved}, we can estimate the HBM bandwidth to be at least $\sim300$GB/sec.

We believe that the efficiency of the computation can be improved by further optimizing the data formatting operations: identifying bottlenecks and rearranging the layout of the tensors. It is also worth noting that the matrix multiplication involves sparse diagonal band kernel with shape of $128\times128$ and we could potentially explore smaller size of kernel to improve efficiency. A possible direction is trying to utilize the convolution operation in TensorFlow to improve the efficiency further. We also expect that as XLA being actively developed and improved over time, it will deliver higher and higher performance.  

\section{Conclusion}
We demonstrate a novel approach to simulate the two-dimensional ferromagnetic Ising model using TensorFlow on Cloud TPU. We adapted the standard checkerboard algorithm to exploit the TPU architecture, in particularly to leverage its efficient matrix operation and the dedicated high-bandwidth low-latency inter-chip interconnect network of the TPU Pod. We calculate the average magnetization and Binder parameter at various temperatures with different lattice sizes. Our numeric estimates on those size-independent quantities match the theoretical results using both float32 and bfloat16 precision. Our benchmarks also demonstrate competitive and linear-scaling performance. The algorithm used in this work can be generalized for three-dimensional Ising model. An interesting direction to follow up would be applying the approach on some of the recent works that push the three-dimensional Ising model simulations to limits, for example\citep{Pushing-To-Limit-3D-2018}.

This work demonstrates how the new Cloud TPU computation resources could be efficiently employed for conventional scientific simulation problems. However, even more significantly, by implementing the entire simulation using TensorFlow framework, we point a direction where the direct integration of machine learning approaches with conventional simulations is possible and could be done easily. For example, the automatic differentiation in TensorFlow\footnote{\raggedright\url{www.tensorflow.org/tutorials/eager/automatic_differentiation}} is readily applicable for optimization of parameters in design problems that use simulations. In the context of this current work, an interesting followup would be finding the optimal $J_{i,j}$ given material properties for the case where $J$ is not uniform across all spin sites. In our view, the research in this direction will bring many interesting advancements and will continue to shape the future of computing.  

\section{Appendices}
\subsection{Proof of Stationarity}
\small
\begin{equation*}
\begin{split}
&\sum_{\pmb{\sigma_w, \sigma_b}}P\{(\pmb{\sigma_w, \sigma_b})\rightarrow(\pmb{\sigma_w^*, \sigma_b^*})\}\pi(\pmb{\sigma_w, \sigma_b}) \\
=&\sum_{\pmb{\sigma_w, \sigma_b}}P\{(\pmb{\sigma_w, \sigma_b})\rightarrow(\pmb{\sigma_w^*, \sigma_b})\}P\{(\pmb{\sigma_w^*, \sigma_b})\rightarrow(\pmb{\sigma_w^*, \sigma_b^*})\}\pi(\pmb{\sigma_w, \sigma_b}) \\
=&\sum_{\pmb{\sigma_w, \sigma_b}}\prod_{i\in w}P(\sigma_i\rightarrow\sigma_i^*|\pmb{\sigma_b})\prod_{i\in b}P(\sigma_i\rightarrow\sigma_i^*|\pmb{\sigma_w^*})\prod_{i\in w}\pi(\sigma_i|\pmb{\sigma_b})\cdot \pi(\pmb{\sigma_b}) \\
=&\sum_{\pmb{\sigma_w, \sigma_b}}\prod_{i\in w}P(\sigma_i\rightarrow\sigma_i^*|\pmb{\sigma_b})\pi(\sigma_i|\pmb{\sigma_b})\prod_{i\in b}P(\sigma_i\rightarrow\sigma_i^*|\pmb{\sigma_w^*})\cdot \pi(\pmb{\sigma_b}) \\
&\textrm{By detailed balance of single spin flip:} \\
& P(\sigma_i\rightarrow\sigma_i^*|\pmb{\sigma_b})\pi(\sigma_i|\pmb{\sigma_b}) = P(\sigma_i^*\rightarrow\sigma_i|\pmb{\sigma_b})\pi(\sigma_i^*|\pmb{\sigma_b}) \\  
=&\sum_{\pmb{\sigma_w, \sigma_b}}\prod_{i\in w}P(\sigma_i^*\rightarrow\sigma_i|\pmb{\sigma_b})\pi(\sigma_i^*|\pmb{\sigma_b})\prod_{i\in b}P(\sigma_i\rightarrow\sigma_i^*|\pmb{\sigma_w^*})\cdot \pi(\pmb{\sigma_b}) \\
=&\sum_{\pmb{\sigma_w, \sigma_b}}\prod_{i\in w}P(\sigma_i^*\rightarrow\sigma_i|\pmb{\sigma_b})\prod_{i\in b}P(\sigma_i\rightarrow\sigma_i^*|\pmb{\sigma_w^*})\prod_{i\in w}\pi(\sigma_i^*|\pmb{\sigma_b})\cdot \pi(\pmb{\sigma_b}) \\
=&\sum_{\pmb{\sigma_w, \sigma_b}}\prod_{i\in w}P(\sigma_i^*\rightarrow\sigma_i|\pmb{\sigma_b})\prod_{i\in b}P(\sigma_i\rightarrow\sigma_i^*|\pmb{\sigma_w^*})\pi(\sigma_i|\pmb{\sigma_w^*})\cdot \pi(\pmb{\sigma_w^*}) \\
& \textrm{By detailed balance of single spin flip:}\\ 
& P(\sigma_i\rightarrow\sigma_i^*|\pmb{\sigma_w^*})\pi(\sigma_i|\pmb{\sigma_w^*})=P(\sigma_i^*\rightarrow\sigma_i|\pmb{\sigma_w^*})\pi(\sigma_i^*|\pmb{\sigma_w^*}) \\
=&\sum_{\pmb{\sigma_w, \sigma_b}}\prod_{i\in w}P(\sigma_i^*\rightarrow\sigma_i|\pmb{\sigma_b})\prod_{i\in b}P(\sigma_i^*\rightarrow\sigma_i|\pmb{\sigma_w^*})\pi(\sigma_i^*|\pmb{\sigma_w^*})\cdot \pi(\pmb{\sigma_w^*}) \\
=& \pi(\pmb{\sigma_w^*, \sigma_b^*})\sum_{\pmb{\sigma_w, \sigma_b}}\prod_{i\in w}P(\sigma_i^*\rightarrow\sigma_i|\pmb{\sigma_b})\prod_{i\in b}P(\sigma_i^*\rightarrow\sigma_i|\pmb{\sigma_w^*}) \\
=& \pi(\pmb{\sigma_w^*, \sigma_b^*})
\end{split}
\end{equation*}
\normalsize

\begin{figure*}[h!]
\centering
\includegraphics[width=0.9\textwidth]{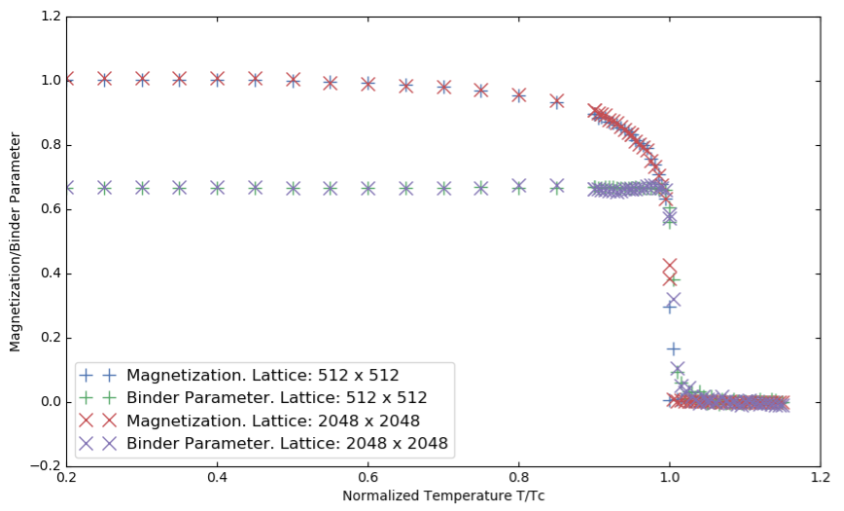}
\caption{Magnetization and Binder parameters simulated using the new algorithm. At lattice size of $[512 \times 512]$, for each data point, we first perform $500,000$ whole-lattice flipping as burn-in and then average the output from the subsequent $1,500,000$ whole-lattice flipping to get the result. For lattice size of $[2048, 2048]$, we have $2,000,000$ whole-lattice flipping as burn-in and the average of the subsequent $6,000,000$ runs are used to generate the data points.}
\label{fig:newmag}
\end{figure*}

\begin{table*}
\centering
\begin{tabular}{|c|c|c|c|c|}
  \hline
  \begin{tabular}{c}
  core \\ topology 
  \end{tabular}&
  \begin{tabular}{c} per-core \\ lattice dimensions \end{tabular} &
  \begin{tabular}{c} whole \\ lattice size
  \end{tabular} & 
  \begin{tabular}{c} time of whole \\ lattice update (ms) \end{tabular} & \begin{tabular}{c} throughput \\ (flips/ns) 
  \end{tabular} \\
  \hline
  \hline
  $[2, 2]$ & & $(128 \times 448)^2$ & 40.78 & 80.64 \\
  $[3, 3]$ & & $(128 \times 672)^2$ & 40.89 & 180.93 \\
  $[4, 4]$ & & $(128 \times 896)^2$ & 40.91 & 321.52 \\
  $[6, 6]$ & & $(128 \times 1344)^2$ & 40.87 & 724.05 \\
  $[8, 8]$ & $[224, 224] \times 128 $ & $(128 \times 1792)^2$ & 41.06 & 1281.47 \\
  $[11, 11]$ & & $(128 \times 2464)^2$ & 41.06 & 2422.60 \\
  $[16, 16]$ & & $(128 \times 3584)^2$ & 41.10 & 5120.02 \\
  $[23, 23]$ & & $(128 \times 5152)^2$ & 41.16 & 10566.16 \\
  $[32, 32]$ & & $(128 \times 7168)^2$ & 41.15 & 20456.20 \\
  $[45, 45]$ & & $(128 \times 10080)^2$ & 41.46 & 40456.29 \\    
  \hline
  \hline
  $[2, 2]$ & & $(128 \times 896)^2$ & 164.08 & 80.17 \\
  $[3, 3]$ & & $(128 \times 1344)^2$ & 164.06 & 180.39 \\
  $[4, 4]$ & & $(128 \times 1792)^2$ & 164.14 & 320.54 \\
  $[6, 6]$ & & $(128 \times 2688)^2$ & 164.22 & 720.85 \\
  $[8, 8]$ & $[448, 448] \times 128 $ & $(128 \times 3584)^2$ & 164.34 & 1280.59 \\
  $[11, 11]$ & & $(128 \times 4928)^2$ & 164.36 & 2420.88 \\
  $[16, 16]$ & & $(128 \times 7168)^2$ & 164.39 & 5120.83 \\
  $[23, 23]$ & & $(128 \times 10304)^2$ & 164.45 & 10577.86 \\
  $[32, 32]$ & & $(128 \times 14336)^2$ & 164.57 & 20460.92 \\
  $[45, 45]$ & & $(128 \times 20160)^2$ & 164.75 & 40418.07 \\  
  \hline
  \hline
  $[2, 4]$ & & $(128 \times 1792)^2$ & 331.80 & 158.57 \\
  $[4, 8]$ & & $(128 \times 3584)^2$ & 332.08 & 633.75 \\
  $[8, 16]$ & $[896, 448] \times 128 $ & $(128 \times 7168)^2$ & 332.45 & 2532.18 \\
  $[16, 32]$ & & $(128 \times 14336)^2$ & 332.72 & 10120.29 \\
  $[32, 64]$ & & $(128 \times 28672)^2$ & 333.36 & 40403.46 \\  
  \hline
\end{tabular}
\caption{Weak scaling performance of the new implementation with TensorFlow r1.15 on TPU v3. We perform tests with three different density settings. From top to bottom section in the table: loose-packed, dense-packed, and superdense-packed. We notice that in all cases the scaling is very much linear, with very small and essentially negligible step time increase as more number of cores are involved.}
\label{table:ncore-weak}
\end{table*}

\begin{figure*}[h!]
\centering
\includegraphics[width=0.9\textwidth]{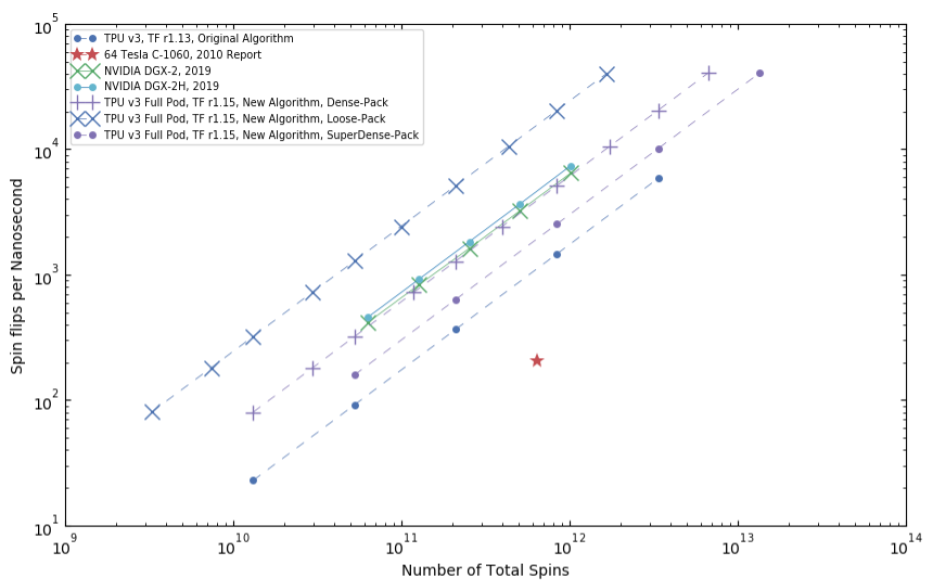}
\caption{Comparison of performance and throughput over various problem sizes. DGX-2 and DGX-2H results are from \citep{Romero2019}.}
\label{fig:comp_all}
\end{figure*}

\begin{table*}
\centering
\begin{tabular}{|c|c|c|c|c|}
  \hline
  \begin{tabular}{c}
  core \\ topology 
  \end{tabular}&
  \begin{tabular}{c} per-core \\ lattice dimensions \end{tabular} &
  \begin{tabular}{c} whole \\ lattice size
  \end{tabular} & 
  \begin{tabular}{c} time of whole \\ lattice update (ms) \end{tabular} & \begin{tabular}{c} throughput \\ (flips/ns) 
  \end{tabular} \\
  \hline
  \hline
  $[2, 4]$ & $[896, 448] \times 128 $ && 330.14 & 159.37 \\
  $[4, 4]$ & $[448, 448] \times 128 $ && 162.55 & 323.67 \\
  $[4, 8]$ & $[448, 224] \times 128 $ && 81.81 & 643.12 \\
  $[8, 8]$ & $[224, 224] \times 128 $ && 41.33 & 1272.94 \\
  $[8, 16]$ & $[224, 112] \times 128 $ & $(128 \times 1792)^2$ & 21.68 & 2427.26 \\
  $[16, 16]$ & $[112, 112] \times 128 $ && 11.08 & 4749.35 \\
  $[16, 32]$ & $[112, 56] \times 128 $ && 6.13 & 8585.73 \\
  $[32, 32]$ & $[56, 56] \times 128 $ && 3.84 & 13704.96 \\
  $[32, 64]$ & $[56, 28] \times 128 $ && 2.86 & 18396.28 \\
  \hline
\end{tabular}
\caption{Strong scaling performance of the new implementations. The performance scales linearly until more than $1,000$ cores are involved in the computation. At which point the communication overhead starts to be a significant part of the run time.}
\label{table:ncore-strong}
\end{table*}

\begin{figure*}[h!]
\centering
\includegraphics[width=0.9\textwidth]{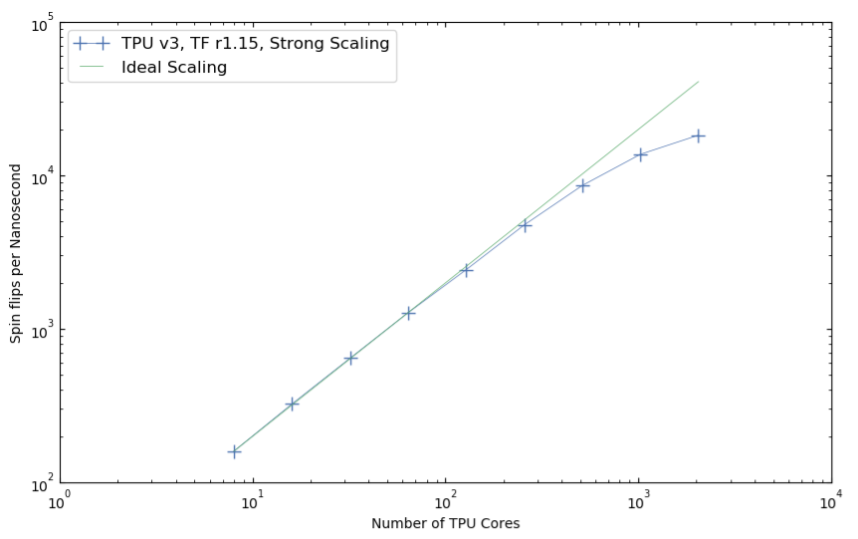}
\caption{The strong scaling performance curve from the new implementation v.s. the ideal linear scaling.}
\label{fig:strong_scaling}
\end{figure*}

\subsection{Further Optimization and Scaling}
In this section, we present additional optimization and further up-scaling of the problem. The additional optimization involves a detailed implementation of the computation of the nearest neighbor energy contribution: \texttt{tf.nn.convol2D} is used, instead of batch multiplication. Detailed implementation is also open-sourced and can be found in \footnote{ \texttt{https://github.com/google-research/google-research/simulation\_research/ising\_model}}. This approach more efficiently leverages MXU's computation power by packing more operations together for each memory load operation. Together with the improvements from the new version of TensorFlow (r1.15), we achieve a $\sim80\%$ performance improvement. In addition, we also utilize all available $2048$ TPU cores in a TPU v3 POD (previously, we only utilized one quarter of a full pod). 

Figure~\ref{fig:newmag} shows the simulated magnetization and Binder parameters using the new implementation and it confirms the new algorithm continues to produce the correct results. 

In our weak-scaling performance tests, we explored different density of work-load: loose-packed ($[224, 224] \times 128$ per core), dense-packed ($[448, 448] \times 128$ per core) and superdense-packed ($[896, 448] \times 128$ per core). We demonstrated that in all cases, they all scale linearly and the largest problem we can handle is $4
\times$ of the largest problem we previously reported. Table~\ref{table:ncore-weak} provides detailed results. We also provide a plot of all available reported performance numbers  in Figure~\ref{fig:comp_all}.

We also perform strong-scaling performance tests. The size of the problem we choose is the $(128 \times 1792)^2$. Table~\ref{table:ncore-strong} shows the results. The scaling stays relatively linear for smaller number of cores, but when more than $1000$ cores are involved, the overhead of communication starts to be a significant part of the run time. From Figure~\ref{fig:strong_scaling} this can also be observed clearly

\section*{Acknowledgements}
We would like to thank Blake Hechtman, Brian Patton, Cliff Young, David Patterson, Naveen Kumar, Norm Jouppi, Rif A. Saurous, Yunxing Dai, Zak Stone, and many more for valuable discussions and helpful comments, which have greatly improved the paper.

\bigskip
\begin{center}
{\large\bf SUPPLEMENTAL MATERIALS}
\end{center}
\textbf{Colaboratory} is hosted on \texttt{https://github.com/google-research/google-research} in the directory of \texttt{simulation\_research/ising\_model}.

\bibliographystyle{plain}

\end{document}